\documentclass[prl,twocolumn,showpacs,preprintnumbers,amsmath,amssymb]{revtex4}

\usepackage{graphicx}
\usepackage{dcolumn}
\usepackage{bm}

\begin{document}


\title{\bf Excitations in one-dimensional $S=1/2$ 
quantum antiferromagnets} 

\author{Kai P. Schmidt and G\"otz S. Uhrig}

\affiliation{Institut f\"ur Theoretische Physik, Universit\"at zu
  K\"oln, Z\"ulpicher Str. 77, D-50937 K\"oln, Germany\\[1mm]
  {\rm(\today)} }

\begin{abstract}
The transition from dimerized to uniform phases is studied in terms of
spectral weights for spin chains using continuous unitary transformations 
(CUTs). The spectral weights in the $S=1$ channel are computed
perturbatively around the limit of strong dimerization. 
We find that the spectral weight is concentrated mainly 
in the subspaces with a small number of elementary triplets (triplons),
even for vanishing dimerization. So, besides spinons,
triplons may  be used as elementary excitations in spin chains. We conclude
that there is no necessity to use fractional excitations in low-dimensional,
undoped or doped quantum antiferromagnets.    
\end{abstract}

\pacs{75.40.Gb, 75.50.Ee, 75.10.Jm} 
\maketitle
\vskip2pc]
The determination of elementary excitations is a fundamental issue in 
condensed matter physics. Especially in the context of high-temperature 
superconductivity, the r\^ole of magnetic excitations in two-dimensional 
quantum antiferromagnets is strongly debated
\cite{laugh97,aeppl99,gruni00,ander00,ho01}. 
The bottom line of the debate is whether or not it is possible 
and/or necessary to consider fractional excitations, so-called spinons with $S=1/2$,
or whether integer excitations, magnons or elementary triplets,
triplons \cite{notiz1}, with $S=1$, can equally be used.
Here we provide clear
indication that even in one-dimensional (1D) spin chains where spinons
are a valid description an alternative description by triplons is possible.
Hence the necessity to resort to fractional excitations is quite
generally questioned.

The spinon concept comes from 1D spin chains ($\delta=0$)
\begin{equation}
 \label{H_DC}
 H' = J_0\sum_i  \left(1+\delta(-1)^i\right) {\bf S}_i{\bf S}_{i+1} 
\end{equation}
 where spinons with $S=1/2$ are established as the elementary excitations
\cite{cloiz62,fadde81}.
 For any finite dimerization $\delta\neq 0$, however,
 the spin chain can be described by massive triplets with $S=1$. 
They can be viewed as bound states of two spinons
\cite{luthe75,halda80,gogol98,uhrig96b} which are, however, not free
due to their confining interaction.
Starting  from strong dimerization, where a description with triplons is certainly 
appropriate, the question arises what happens to the triplons 
for vanishing dimerization. Besides its academic interest,
this question is of great practical importance since in recent years
triplon-based techniques have been advanced considerably
\cite{knett01b,uhrig98c,knett00a,knett00b,trebs00,zheng01a,schmi01}.
Our recent progress has rendered even the computation of 
spectral \emph{densities} possible 
\cite{knett01b,schmi01,windt01,gruni02b,knett02}.

In this work, we investigate  to which extent a triplon-based
description is possible by computing spectral weights of multi-triplon
contributions for dimerized spin chains. Two scenarios are conceivable
for $\delta\to 0$:\\
(i) The spectral weight 
is distributed rather evenly over the multi-triplon channels
implying that the weight of a particular channel
is small and that a large number of channels 
must be taken into account, see discussion in
Ref.~\onlinecite{uhrig96b}. This scenario would
make an approach in terms of triplons difficult to use, hence
inappropriate.\\
(ii) The spectral weight is found mainly in the channels with a \emph{small}
 number of triplons, implying that a triplon approach is very useful and
appropriate because spectral properties
can be computed from the dynamics of a small number of
excitations in a small number of 
channels.

\paragraph{Scenario (i)}
Let us look at a field theoretic description for indications for 
this scenario. Using abelian bosonization 
\cite{halda80,halda81b,affle90,voit95} the low-lying states
of an anisotropic spin chain are described by the Hamiltonian
\begin{equation}
\label{H_FB}
H_{\rm FT} = \frac{v}{2\pi}\int_{-\infty}^\infty 
\left[  K (\pi\Pi(x))^2 + K^{-1} (\partial_x\Phi(x))^2 \right] dx 
\end{equation}
where $v\propto J_0$ is the spin-wave velocity and
$K$ is the interaction parameter, e.g.\  $K=1$ 
 for the XY-model, equivalent to free fermions, and
$K=1/2$ for the isotropic case. A local operator is
\begin{eqnarray}
S^z_j &= & (2\pi)^{-1} \partial_x \Phi(j) + A (-1)^j \cos(2\Phi(j))
\label{composite}
\end{eqnarray}
where the lattice constant is unity, and $A$  a non-universal
constant. The undimerized Hamiltonian (\ref{H_DC}) corresponds essentially
to free bosons (\ref{H_FB}).
A finite dimerization is accounted for by an additional term
proportional to $\delta \int_{-\infty}^\infty \cos(2\Phi(x))dx$
leading to a sine-Gordon model. For $K<2$ the system is massive with a
 gap $\Delta\propto\delta^{1/(2-K)}$ 
($=\delta^{2/3}$ for the isotropic chain). 

A single bosonic mode of (\ref{H_FB}) is created by
\begin{equation}
\label{bosons}
b_k^\dagger = 1/\sqrt{2}\left(
\tilde\Phi(k)/N_k - i N_k \tilde\Pi(k)
\right)
\end{equation}
where $N_k = \sqrt{\pi K/|k|}$.
The quantities with tilde are the Fourier transforms of
the real-space fields. We expand
the excited state $S^z_j |0\rangle$ in states of various number
of bosons with focus on the vicinity of momentum
$\pi$ where most of the weight in the dynamic structure factor
is found: $S^z_j\propto \cos(2\Phi(j))$. 
The coefficients of one- and two-boson states are
\begin{subequations}
\begin{eqnarray}
c_k &=& \langle 0|b_k \cos(2\Phi)|0\rangle = 
\langle 0|[b_k, \cos(2\Phi)]|0\rangle\\
c_{k,q} &=& \langle 0|b_k b_q \cos(2\Phi)|0\rangle
= \langle 0|[b_k [b_q ,\cos(2\Phi)]]|0\rangle\ .
\end{eqnarray}
\end{subequations}
We make use of
\begin{equation}
[b_k,2\Phi(j)]=\sqrt{{K}/{|k|}}\sqrt{{2\pi}/{L}}\exp(ikj)
\end{equation}
where the momenta are discretized by a finite system size $L$ 
to ensure normalizability. This yields $c_k = 0$ and
\begin{eqnarray}
c_{k,q} &\propto& 
\textstyle\frac{K}{\sqrt{|k||q|}}{\frac{2\pi}{L}}\exp(i(k+q)j) \Delta^K
\end{eqnarray}
where $\langle \sin(2\Phi) \rangle =0 $ and 
$\langle \cos(2\Phi) \rangle \propto \Delta^K$ is used. So the total weight
$W^z_1$ in the one-boson channel is always zero.
The total weight $W_2^z$ in the two-boson channel is
\begin{eqnarray}\nonumber
W^z_2 &\propto& \textstyle\sum_{k,q} |c_{k,q}|^2
 \propto  \Delta^{2K} \int_{v|k|,v|q| >\Delta} \frac{K^2}{{|k||q|}} dk dq\\
\nonumber
& \propto & \Delta^{2K} \ln(\Delta)^2\\
& \propto & \delta^{2K/(2-K)} \ln(\delta)^2
\end{eqnarray}
in leading order in $\ln(\delta)$. Generally, all channels with an odd
number of bosons carry no weight (at momentum $\pi$) whereas channels
with $2n$ bosons have 
\begin{equation}
\label{decay1}
W^z_{2n} \propto \delta^{2K/(2-K)}\ln(\delta)^{2n}\ .
\end{equation}
whence we conclude that \emph{any} single channel becomes 
\emph{irrelevant} on 
$\delta\to0$. Only the consideration of an infinite number allows
to treat vanishing dimerization correctly.
This appears to be sound evidence for scenario (i).

\paragraph{Scenario (ii)}
Our main argument is an explicit calculation of spectral weights
for the first triplon channels. First, however, we  challenge the \emph{general}
validity of the previous argument by four basic considerations.

(a) Considering free fermions ($K=1$) we know that the operator 
$S^z_j=n_j-1/2$ excites a particle-hole continuum at all wave-vectors.
So the dynamic correlations of this operator are 
exhaustively described by \emph{two} elementary excitations, a
particle and a hole, independent from the dimerization $\delta$.

(b) Considering the isotropic spin chain ($K=1/2$) it was shown
that $72.89$ \% of the total weight (sum over all wave vectors)
 of the dynamic structure factor 
is described by the \emph{two}-spinon continuum \cite{karba97}.

(c) It is undisputed that the dimerized isotropic spin chain
displays a \emph{single}-mode peak with finite spectral weight at all
wave vectors
as long as $\delta\neq 0$ \cite{gogol98,uhrig96b}.

(d) By construction, the bosonic modes of the field theory
(\ref{bosons}) exist at \emph{small}
 momenta $k\approx 0$ \cite{halda81b,voit95,luthe74b}. The dynamics at momenta
close to $\pi$ is captured by the superposition of an infinite
number of these modes as becomes evident from Eq.~(\ref{composite})
and from Eq.~(\ref{decay1}).

The above arguments prove that there can be 
several, rather different looking descriptions for the \emph{same} physics.
One may \emph{not} conclude
from the validity of a particular description, e.g.\ in terms of 
a multitude of modes, that another description, e.g.\ in terms of
only a few modes, is not valid.

Bearing these considerations in mind, we turn to a continuous
unitary transformation (CUT) of the Hamiltonian (\ref{H_DC}) onto an
effective Hamiltonian in terms of triplons. To proceed
 perturbatively  we transform (\ref{H_DC}) into
\begin{equation}
 \label{H_TDC}
 H = H'/J =  \sum_i \left[ {\bf S}_{2i}{\bf S}_{2i+1} + 
\lambda{\bf S}_{2i}{\bf S}_{2i-1} \right]\ ,
\end{equation}
where $J=J_0(1+\delta)$ and $\lambda=(1-\delta)/(1+\delta)$.
We expand about isolated dimers at $\lambda=0$. 
A CUT is used to map the Hamiltonian $H$ to
 an effective Hamiltonian $H_{\rm eff}$ which conserves the number of triplets
 on the strong bonds, i.e.\ $[H_{\rm 0},H_{\rm eff}]=0$ where 
$H_{\rm 0}:=H|_{\lambda=0}$ \cite{knett00a}. Hence
these triplets become the elementary excitations, triplons, of the
effective Hamiltonian.
The ground state of $H_{\rm eff}$ is the triplon vacuum. 
An infinitesimal antihermitian generator $\eta$ induces the flow
${dH}/{dl} = [\eta (l),H(l)]$ \cite{wegne94},
where  $l$ is an auxiliary variable, 
$H(0) = H$, $H(\infty) = H_{\rm eff}$. An appropriate choice for $\eta$ is
\begin{equation}
 \label{Generator}
 \eta_{i,j}(l) = {\rm sgn} ([H_{0}]_{i,i}-[H_{0}]_{j,j})H_{i,j}(l)
\end{equation}
where the matrix elements $\eta_{i,j}$ and $H_{i,j}$ are given in an 
eigen-basis of $H_{\rm 0}$. 
The choice (\ref{Generator}) retains only triplon conserving processes
eliminating all parts of $H$ changing the number of triplons \cite{knett00a}. 

In order to determine spectral weights corresponding to a given
observable $R$ this observable is transformed in the same way as
the Hamiltonian
\begin{equation}
 \label{CUT-O}
 {dR}/{dl} = [\eta (l),R(l)]
\end{equation} 
leading for $l\to \infty$ to the effective observable $R_{\rm eff}$.
After the transformation, i.e.\ at $l=\infty$,
the subspaces with different number of triplons are disentangled:
the Hamiltonian conserves the number of triplons and does not
mix subspaces with different numbers of triplons.
Given this property,
it makes sense to define the spectral weight of each of these subspaces
by
\begin{equation}
I_n = \langle 0 |R^{\dagger}_{\rm eff} P_n  
R_{\rm eff}^{\phantom\dagger}|0\rangle
\qquad n\in\{0,1,2,3,\ldots\}
\end{equation}
where $P_n$ projects onto the subspace with $n$ triplons
\cite{knett01b,schmi01,windt01,gruni02b}.
This requires only explicit counting since the
ground state $|0\rangle$ after the transformation is the triplon 
vacuum, i.e.\ it is the product state of singlets on all
strong bonds.  The
technically involved explicit definition of ``reduced exclusive matrix
elements'' is avoided \cite{barne99}. Note that the definition of
spectral weights like $I_n$ is useless if the
triplon number is not conserved. Only the triplon conservation ensures
that the total correlation splits into additive contributions
from the various subspaces.

The CUT cannot be carried out completely. Perturbatively
the differential equations are truncated when terms of a certain order
in  $\lambda$ arise \cite{knett00a}. 
The local spin component $S^z_j$ is the observable $R$.
The effective observable $R_{\rm eff}$ is calculated till order $7$ in the 
one-, two-, and three-triplon sector. The four-triplon sector
is transformed till order $6$. 

The total weight $I_{\rm tot}=\sum_0^\infty I_n$ 
is given by the  sum rule
$I_{\rm tot} = \langle R^\dagger R \rangle = \langle (S^z_j)^2 \rangle =1/4$
which serves as sensitive check for the validity of the 
results. In the following we will discuss
the relative weights $I_{n,{\rm rel}}=I_n/I_{\rm tot}= 4I_n$.

\begin{table}[th]
\begin{tabular}{|c||c|c|c|}
 dlogPad\'{e} &a) Zero $\lambda_0$ &b) $\gamma|_{\lambda=\lambda_0}$ & c)
$\gamma|_{\lambda=1}$ \\
 \hline
 $[4,2]$ &           &           & ${ 0.32524}$  \\
 $[3,3]$ & $1.02503$ & $0.36798$ & ${ 0.32891}$  \\
 $[2,4]$ & $1.09817$ & $0.58184$ & ${ 0.34110}$  \\
 $[1,5]$ & $1.09817$ & $0.58184$ & ${ 0.31457}$  \\
 $[0,6]$ &           &           & ${ 0.31458}$  
\end{tabular}
\caption{Relative weight $I_{\rm 1,rel}$ of the one-triplon channel. 
a) position of the singularity  from unbiased
approximants; b) exponent at the unbiased positions;
c) exponents in the biased approximants.
(blanks: approximants without singularity).
\label{tab1}
}
\end{table}
For isolated dimers ($\lambda=0$), the total
 spectral weight lies in the one-triplon channel $I_{\rm 1,rel}=1$. As 
$\lambda$ increases $I_{\rm 1,rel}$ decreases and the weight flows into the
multi-triplon sectors. In Tab.~\ref{tab1}, the results for unbiased
dlogPad\'{e} approximants indicate a singularity at
$\lambda\approx 1$. From the physics of the Hamiltonian (\ref{H_DC})
we know that the singularity is located at $\lambda=1$ where the system
becomes critical. Thus it is advised to investigate
approximants biased to display the singularity at unity:  
$I_1 \propto (1-\lambda)^\gamma$
(last column in Tab.~\ref{tab1}). 
The exponent is found to be $\gamma =0.325\pm 0.016$.
which leads us to conjecture that it takes
exactly the value $\gamma=1/3$. More generally, we presume that any 
single mode, which vanishes due to mixing with
 a continuum with square-root singularities at the
band edges, looses its weight like $I_{\rm mode} \propto \sqrt{\Delta \omega}$,
 where $\Delta \omega$ is the
distance of the mode to the band edge of the continuum. The weight in the
continuum is assumed to be constant.

We support our claim by considering
a generic resolvent $I(\omega)= 1/(\omega - a - \Sigma(\omega))$
with $\Sigma(\omega) = [\omega+\sqrt{\omega^2 -4}]/2$ for $\omega\le -2$. 
Such a resolvent
appears for instance in the dynamics of two hard-core particles hopping
from site to site in one dimension  at given
total momentum \cite{uhrig96b}. The constant $a$ allows to tune
a nearest-neighbor interaction ($a<0$: attraction, $a>0$: repulsion)
whereas $\Sigma(\omega)$ incorporates the kinetic energy of the relative
motion. For $a\le -1$ a bound state emerges from the continuum. It is
separated from the continuum by the energy $\Delta \omega = -(2+a+1/a)$.
Its weight $I_{\rm mode}$
is  $1/(1-\partial_\omega\Sigma(-2-\Delta \omega))$ implying
$I_{\rm mode}\propto \sqrt{\Delta\omega}$ for $ \Delta\omega \to 0$.

For dimerized spin chains, the single triplon mode
is separated from the continuum by an energy $\Delta\omega$
of the order of the energy gap $\Delta$ \cite{uhrig96b}.  
Using $I_{\rm mode} \propto \sqrt{\Delta \omega}$ for each
total momentum and integrating then over all momenta to 
obtain the local weights
we find that the single mode looses its weight as
$\sqrt{\Delta}\propto \delta^{1/3}$ which agrees excellently with
the extrapolations. 

\begin{table}[th]
\begin{tabular}{|c||c|c|c|c|}
 dlogPad\'{e} & a) Zero $\lambda_0$ & b) $\gamma_2|_{\lambda=\lambda_0}$ & 
 c) $\gamma_2|_{\lambda=1}$  & d) $I_{2,{\rm rel}}$\\
 \hline
 $[5,0]$ &           &            &  $\star$  & $1.0618$\\
 $[4,1]$ &           &            &  $\star$  & $0.9818$\\
 $[3,2]$ &           &            & $-0.7601$ & $0.9976$\\
 $[2,3]$ & $0.9908$  & $-0.7323$  & $-0.7603$ & $\star$ \\
 $[1,4]$ &           &            &  $\star$  & $0.9895$
\end{tabular}
\caption{Relative weight $I_{\rm 2,rel}$ of the two-triplon channel. 
a) position of the singularity from unbiased
approximants for $\partial_\lambda I_{\rm 2,rel}$; 
b) exponent at the unbiased positions;
c) exponents in the biased approximants;
d) $I_{\rm 2,rel}|_{\lambda=1}$  integrated
from the biased (position $\lambda=1$ and exponent
$-2/3$) approximants
(approximants  without singularity: blanks;  with spurious poles: stars).
\label{tab2}}
\end{table}
In the two-triplon channel, there is no indication for a zero of $I_2$ at $\lambda=1$. 
On the contrary, Pad\'e approximants indicate significant weight at criticality.
The weight in the two-triplon channel is the weight transferred from the
one-triplon channel minus the weight transferred further to  channels
with three and more triplons. Hence it is natural to assume the existence
of a singularity with exponent 1/3 in $I_2$. 
But if this singularity is not linked to a zero,
dlogPad\'e approximants cannot detect it. Hence we investigate
the derivative $\partial_\lambda I_2$ which should be governed by
a divergence with exponent -2/3. Indeed, Pad\'e (not shown) and 
dlogPad\'e approximants (Tab.~\ref{tab2}) indicate a singularity
at $\lambda\approx 1$. Approximants biased
to a singularity of $\partial_\lambda I_2$
at $\lambda=1$ yield exponents $\gamma_2\approx-0.76$.
The corresponding value of $I_{2,{\rm rel}}$ found from integrating
$\partial_\lambda I_2$ is 1.25. Since this value overestimates 
the sum rule by at least 25\%, we conclude that the exponent 
$\gamma_2\approx-0.76$ is too large in absolute value. Thus,  we 
bias the approximants to the expected behavior 
$\partial_\lambda I_2 \propto (1-\lambda)^{-2/3}$. 
In the last column of Tab.~\ref{tab2}
the ensuing values for $I_{2,{\rm rel}}$ are given. 
Quite unexpectedly, the results conclusively
point to a spectral weight very close to unity! Since diagonal approximants
usually yield the most reliable results
we retain the value $I_{2,{\rm rel}}\approx 0.998$, keeping a possible error
of a few percent in mind.

The sum rule corroborates the above result strongly. Pad\'e 
and unbiased dlogPad\'e approximands consistently show that
$I_{3,{\rm rel}}$ is not larger than about  $3\cdot 10^{-4}$
which   agrees perfectly with the value close to unity for $I_{2,{\rm rel}}$.
The inclusion of a singularity does not enhance
$I_{3,{\rm rel}}$. The 
 biased approximants for $\partial_\lambda I_3$
(position $\lambda=1$ and exponent $-2/3$) 
yield also only a contribution of about $2\cdot 10^{-4}$ ([2,2]).
Finally, the Pad\'e and dlogPad\'e approximants
for the four-triplon contribution $I_{4,{\rm rel}}$
consistently indicate values well below $10^{-4}$.
The biased approximant for $\partial_\lambda I_4$
(position $\lambda=1$ and exponent $-2/3$) even
yield values below  $10^{-5}$. Therefore, we
conclude that the contributions of channels with four and
more triplons can be safely neglected.

Fig.~1 shows the final results for $I_{1,{\rm rel}}, I_{2,{\rm rel}}$
and $I_{3,{\rm rel}}$; the tiny four- and more triplon
 contributions are neglected. 
The sum rule is excellently fulfilled to within $\approx 0.003$
for all values of $\lambda$ supporting the above analysis. 
\begin{figure}[htbp]
  \begin{center}
    \includegraphics[width=\columnwidth]{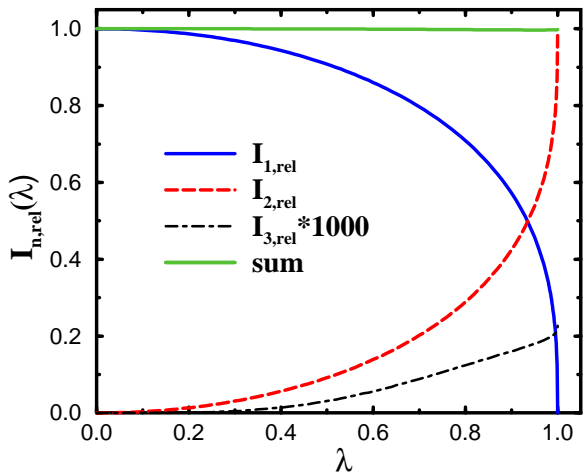} 
    \caption{Relative spectral weights $I_{\rm n,rel}(\lambda)$ of the dynamic
      structure factor in the dimerized chain. Depicted are biased 
      approximants with singularity at $\lambda=1$ and exponent $1/3$ 
      for $I_{\rm 1,rel}$ ([4,2], black solid line);  exponent $-2/3$ 
      in the derivative of $I_{\rm 2,rel}$ ([3,2], dashed line) and similarly 
      for $1000\cdot I_{\rm 3,rel}$ ([2,2], dashed-dotted line). The grey line
      is the sum $\sum_{n=1}^3 I_{\rm n,rel}$.} 
    \label{fig:Spectral_Weights_S1}
  \end{center}
\end{figure}

In conclusion, we have found that even the dynamic structure factor of
the critical uniform spin chain can be
described to about 99\% by two elementary triplets (triplons).
This shows that scenario (ii) is correct for spin chains.
A description  in terms of a few triplons is possible. This lays the
foundation for a new route to spectral densities in many
 models (first results in \cite{knett01b,zheng03a}).

Since the spin chain is the archetype of a gapless critical model
described by spinons we conclude that neither the occurrence of 
fractional excitations nor the vanishing of the gap precludes
the applicability of an approach in terms of integer triplons.
Even a larger part of the spectral weight is covered by two-triplon
states than is covered by two-spinon states (72.89\%) \cite{karba97},
which calls for further investigations of the relation
between spinon and transformed triplon states.
Since these results hold in spin chains, the home field of spinons,
we conclude that a large class of low-dimensional quantum antiferromagnets
is accessible by calculations based on integer excitations. 
If this is true for undoped antiferromagnets there is no necessity either
to resort to fractional excitations for doped antiferromagnets. 
Hence our finding embodies an important message for
potential theories for high-temperature superconductors.

We thank M.~Gr\"uninger and E.\ M\"uller-Hartmann for
 stimulating and encouraging discussions 
 and the DFG  for financial support in SP 1073 and in SFB 608. 


\begin{thebibliography}{10}

\bibitem{laugh97}
R.~B. Laughlin, Phys. Rev. Lett. {\bf 79},  1726  (1997).

\bibitem{aeppl99}
G. Aeppli {\it et~al.}, phys. stat. sol. (b) {\bf 215},  519  (1999).

\bibitem{gruni00}
M. Gr\"uninger {\it et~al.}, Phys. Rev. B {\bf 62},  12422  (2000).

\bibitem{ander00}
P.~W. Anderson, Science {\bf 288},  480  (2000).

\bibitem{ho01}
Ch.-M.Ho, V.~N. Muthukumar, M. Ogata, and P.~W. Anderson, Phys. Rev. Lett. {\bf
  86},  1626  (2001).

\bibitem{notiz1}
Previously, we used the term 
``elementary triplets'' to distinguish 
three-fold degenerate elementary excitations from
magnons which are the elementary excitations
of long-range \emph{ordered} magnets.  We henceforth 
use the term ``triplon'' instead to have a
shorter expression distinguishing  more clearly
from composite triplets.

\bibitem{cloiz62}
J. des Cloizeaux and J.~J. Pearson, Phys. Rev. {\bf 128},  2131  (1962).

\bibitem{fadde81}
L.~D. Faddeev and L.~A. Takhtajan, Phys. Lett. {\bf 85A},  375  (1981).

\bibitem{luthe75}
A. Luther and I. Peschel, Phys. Rev. B {\bf 12},  3908  (1975).

\bibitem{halda80}
F.~D.~M. Haldane, Phys. Rev. Lett. {\bf 45},  1358  (1980).

\bibitem{gogol98}
A.~O. Gogolin, A.~A. Nersesyan, and A.~M. Tsvelik,
{\em Bosonization and Strongly Correlated Systems}
(Cambridge University Press, Cambridge, 1998)

\bibitem{uhrig96b}
G.~S. Uhrig and H.~J. Schulz, Phys. Rev. B {\bf 54},  R9624  (1996);
erratum {\bf 58},  2900  (1998).

\bibitem{knett01b}
C. Knetter, K.~P. Schmidt, M. Gr\"uninger, and G.~S. Uhrig, Phys. Rev. Lett.
{\bf 87},  167204  (2001).

\bibitem{uhrig98c}
G.~S. Uhrig and B. Normand, Phys. Rev. B {\bf 58},  R14705  (1998).

\bibitem{knett00a}
C. Knetter and G.~S. Uhrig, Eur. Phys. J. B {\bf 13},  209  (2000).

\bibitem{knett00b}
C. Knetter, A. B\"uhler, E. M\"uller-Hartmann, and G.~S. Uhrig, Phys. Rev.
  Lett. {\bf 85},  3958  (2000).

\bibitem{trebs00}
S. Trebst {\it et~al.}, Phys. Rev. Lett. {\bf 85},  4373  (2000).

\bibitem{zheng01a}
W. Zheng {\it et~al.}, Phys. Rev. B {\bf 63},  144410 and 144411  (2001).

\bibitem{schmi01}
K.~P. Schmidt, C. Knetter, and G.~S. Uhrig, Europhys. Lett. {\bf 56},  877
  (2001).

\bibitem{windt01}
M. Windt {\it et~al.}, Phys. Rev. Lett. {\bf 87},  127002  (2001).

\bibitem{gruni02b}
M. Gr\"uninger {\it et~al.}, J. Phys. Chem. Solids {\bf 63},  2335  (2002).

\bibitem{knett02}
C. Knetter, K.~P. Schmidt, and G.~S. Uhrig, Physica B {\bf 312-313},  527
  (2002).

\bibitem{halda81b}
F.~D.~M. Haldane, J. Phys. {\bf C14},  2585  (1981).

\bibitem{affle90}
I. Affleck,  in {\em Fields, Strings and Critical Phenomena} (Elsevier,
  North-Holland, Amsterdam, 1990), p.\ 566.

\bibitem{voit95}
J. Voit, Rep. Prog. Phys. {\bf 58},  977  (1995).

\bibitem{karba97}
M. Karbach {\it et~al.}, Phys. Rev. B {\bf 55},  12510  (1997).

\bibitem{luthe74b}
A. Luther and I. Peschel, Phys. Rev. B {\bf 9},  2911  (1974).

\bibitem{wegne94}
F.~J. Wegner, Ann. Physik {\bf 3},  77  (1994).

\bibitem{barne99}
T. Barnes, J. Riera, and D.~A. Tennant, Phys. Rev. B {\bf 59},  11384  (1999).

\bibitem{zheng03a}
W. Zheng, C.~J. Hamer, and R.~R.~P. Singh, cond-mat/0211346.

\end{thebibliography}


\end{document}